\documentstyle[12pt,prd,aps,epsfig,preprint]{revtex}

\begin{document}
\title{{\bf Bound-States of the Spinless Salpeter Equation for the }${\cal PT}${\bf %
-Symmetric Generalized Hulth\'{e}n Potential by the Nikiforov-Uvarov Method }}
\author{Sameer M. Ikhdair\thanks{%
sameer@neu.edu.tr} and \ Ramazan Sever\thanks{%
sever@metu.edu.tr}}
\address{$^{\ast }$Department of Physics, \ Near East University, Nicosia, North\\
Cyprus, Mersin 10, Turkey\\
$^{\dagger }$Department of Physics, Middle East Technical University, 06531\\
Ankara, Turkey.}
\date{\today
}
\maketitle

\begin{abstract}
The one-dimensional spinless Salpeter equation has been solved for the $%
{\cal PT}$-symmetric generalized Hulth\'{e}n potential. The Nikiforov-Uvarov
(${\rm NU}$) method which is based on solving the second-order linear
differential equations by reduction to a generalized equation of
hypergeometric type is used to obtain exact energy eigenvalues and
corresponding eigenfunctions. We have investigated the positive and negative
exact bound states of the ${\rm s}$-states for different types of complex
generalized Hulth\'{e}n potentials.

Keywords: Bethe-Salpeter equation, Energy Eigenvalues and Eigenfunctions;
Generalized Hulth\'{e}n potential; ${\cal PT}$-symmetry, ${\rm NU}$ Method.

PACS numbers: 03.65.-w; 03.65.Fd; 03.65.Ge.
\end{abstract}


\section{Introduction}

\noindent In the past few years there has been considerable work on
non-Hermitian Hamiltonians. Among this kind of Hamiltonians, much attention
has been focused on the investigation of properties of so-called ${\cal PT}%
{\rm -}$symmetric Hamiltonians. Following the early studies of Bender {\it %
et al}. [1], the ${\cal PT}{\rm -}$symmetry formulation has been successfuly
utilized by many authors [2-8]. The ${\cal PT}{\rm -}$symmetric but
non-Hermitian Hamiltonians have real spectra whether the Hamiltonians are
Hermitian or not. Non-Hermitian Hamiltonians with real or complex spectra
have also been analyzed by using different methods [3-6,9]. Non-Hermitian
but ${\cal PT}{\rm -}$symmetric models have applications in different
fields, such as optics [10], nuclear physics [11], condensed matter [12],
quantum field theory [13] and population biology [14].

Exact solution of Schr\"{o}dinger equation for central potentials has
generated much interest in recent years. So far, some of these potentials
are the parabolic type potential [15], the Eckart potential [16,17], the
Fermi-step potential [16,17], the Rosen-Morse potential [18], the Ginocchio
barrier [19], the Scarf barriers [20], the Morse potential [21] and a
potential which interpolates between Morse and Eckart barriers [22]. Many
authors have studied on exponential type potentials [23,24,25,26] and quasi
exactly solvable quadratic potentials [27,28,29]. In addition,
Schr\"{o}dinger, Dirac, Klein-Gordon, and Duffin-Kemmer-Petiau equations for
a Coulomb type potential are solved by using different method
[30,31,32,33,34]. The exact solutions for these models have been obtained
analytically.

Further, using the quantization of the boundary condition of the states at
the origin, Znojil [35] studied another generalized Hulth\'{e}n and other
exponential potentials in non-relativistic and relativistic regions.
Domingues-Adame [36] and Chetouani {\it et al.} [37] also studied
relativistic bound states of the standard Hulth\'{e}n potential. On the
other hand, Rao and Kagali [38] investigated the relativistic bound states
of the exponential-type screened Coulomb potential by means of the
one-dimensional Klein-Gordon equation. However, it is well-known that for
the exponential-type screened Coulomb potential there is no explicit form of
the energy expression of bound-states for Schr\"{o}dinger [39], KG [38] and
also Dirac [16] equations. In a recent work [31], \c{S}im\c{s}ek and E\u{g}%
rifes have presented the bound-state solutions of one-dimensional ($1D$)
Klein-Gordon equation for ${\cal PT}{\rm -}$symmetric potentials with real
and complex forms of the generalized Hulth\'{e}n potential. In a latter
study [32], E\u{g}rifes and Sever investigated the bound-state solutions of
the $1D$ Dirac equation for real and complex forms of generalized
Hulth\'{e}n potential for ${\cal PT}{\rm -}$symmetric potentials with
complex generalized Hulth\'{e}n potential. In recent works, we have solved
the $1D$ Schr\"{o}dinger equation with the ${\cal PT}{\rm -}$symmetric
modified Hulth\'{e}n and Woods-Saxon (WS) potentials for $\ell \neq 0$
bound-state spectra and their corresponding wave functions [40,41] using the
Nikiforov-Uvarov (NU) method [42]. In the latter case, we investigated the $%
{\cal PT}{\rm -}$symmetric property and the reality of the spectrum for
different real and complex versions of \ the modified WS potentials [41].

In the present study we investigate the bound-states solutions of the $1D$
spinless Salpeter (SS) equation for the real and complex forms of the
generalized Hulth\'{e}n potential and correponding eigen functions using the
same method. Thus, the objective of this work is to deal with the ${\cal PT}%
{\rm -}$symmetric property and the existence of bound-states (i.e., the
reality of energy spectrum) when we solve the spinless Salpeter (SS)
equation for some real and complex potentials with standard generalized
forms. In view of the ${\cal PT}{\rm -}$symmetric formulation, we shall
apply the NU method to solve the $s$-wave SS equation. In this regard, it is
possible to present the theory of special functions by starting from a
differential equation which is based on solving the SS equation by reducing
it into a generalized equation of hypergeometric form. We also seek to
present exact bound states for a family of exponential-type potentials,
i.e., generalized Hulth\'{e}n potential which is reducible to the standard
Hulth\'{e}n potential, Woods-Saxon potential and exponential-type screened
Coulomb potential. This family of potentials have been applied with success
to a number of different fields of physical systems.

The organization of the present work is as follows. After a brief
introductory discussion of the NU method in Section \ref{TNU}, we obtain the
bound-state energy spectra for real and complex cases of generalized
Hulth\'{e}n potentials and their corresponding eigenfunctions in Section \ref
{RSS}. Finally, we end with some results and conclusions in Section \ref{RAC}%
.

\section{The Nikiforov-Uvarov Method}

\label{TNU}In this section we outline the basic formulations of the method.
The Schr\"{o}dinger equation and other Schr\"{o}dinger-type equations can be
solved by using the Nikiforov-Uvarov (${\rm NU}$) method which is based on
the solutions of general second-order linear differential equation with
special orthogonal functions [42]. It is well known that for any given $1D$
radial potential, the Schr\"{o}dinger equation can be reduced to a
generalized equation of hypergeometric type with an appropriate
transformation and it can be written in the following form

\begin{equation}
\psi _{n}^{\prime \prime }(s)+\frac{\widetilde{\tau }(s)}{\sigma (s)}\psi
_{n}^{\prime }(s)+\frac{\widetilde{\sigma }(s)}{\sigma ^{2}(s)}\psi
_{n}(s)=0,
\end{equation}
where $\sigma (s)$ and $\widetilde{\sigma }(s)$ are polynomials, at most of
second-degree, and $\widetilde{\tau }(s)$ is of a first-degree polynomial.
To find particular solution of Eq.(1) we apply the method of separation of
variables using the transformation

\begin{equation}
\psi _{n}(s)=\phi _{n}(s)y_{n}(s),
\end{equation}
which reduces equation (1) into a hypergeometric-type equation

\begin{equation}
\sigma (s)y_{n}^{\prime \prime }(s)+\tau (s)y_{n}^{\prime }(s)+\lambda
y_{n}(s)=0,
\end{equation}
whose polynomial solutions $y_{n}(s)$ of the hypergeometric type function
are given by Rodrigues relation

\begin{equation}
y_{n}(s)=\frac{B_{n}}{\rho (s)}\frac{d^{n}}{ds^{n}}\left[ \sigma ^{n}(s)\rho
(s)\right] ,\text{ \ \ \ }\left( n=0,1,2,...\right)
\end{equation}
where $B_{n}$ is a normalizing constant and $\rho (s)$ is the weight
function satisfying the condition [42]

\begin{equation}
\frac{d}{ds}w(s)=\frac{\tau (s)}{\sigma (s)}w(s),
\end{equation}
where $w(s)=\sigma (s)\rho (s).$ On the other hand, the function $\phi (s)$
satisfies the condition

\begin{equation}
\frac{d}{ds}\phi (s)=\frac{\pi (s)}{\sigma (s)}\phi (s),
\end{equation}
where the linear polynomial $\pi (s)$ is given by
\begin{equation}
\pi (s)=\frac{\sigma ^{\prime }(s)-\widetilde{\tau }(s)}{2}\pm \sqrt{\left(
\frac{\sigma ^{\prime }(s)-\widetilde{\tau }(s)}{2}\right) ^{2}-\widetilde{%
\sigma }(s)+k\sigma (s)},
\end{equation}
from which the root $k$ is the essential point in the calculation of $\pi
(s) $ is determined$.$ Further, the parameter $\lambda $ required for this
method is defined as
\begin{equation}
\lambda =k+\pi ^{\prime }(s).
\end{equation}
Further, in order to find the value of $k,$ the discriminant under the
square root is being set equal to zero and the resulting second-order
polynomial has to be solved for its roots $k_{+,-}$. Thus, a new eigenvalue
equation for the ${\rm SE}$ becomes

\begin{equation}
\lambda _{n}+n\tau ^{\prime }(s)+\frac{n\left( n-1\right) }{2}\sigma
^{\prime \prime }(s)=0,\text{ \ \ \ \ \ \ }\left( n=0,1,2,...\right)
\end{equation}
where

\begin{equation}
\tau (s)=\widetilde{\tau }(s)+2\pi (s),
\end{equation}
and it must have a negative derivative.

\section{Radial Spinless Salpeter Equation}

\label{RSS}The relativistic wave Salpeter equation [43] is constructed by
considering the kinetic energies of the constituents and the interaction
potential. In addition, the spinless Salpeter (SS) for the case of two
particles with unequal masses $m_{1}$ and $m_{2},$ interacting by a
spherically symmetric potential $V(r)$ in the center-of-momentum system of
the two particles is given by

\begin{equation}
\left[
\mathrel{\mathop{\sum }\limits_{i=1,2}}%
\sqrt{-\Delta +m_{i}^{2}}+V(r)-M\right] \chi \left( {\bf r}\right) =0,\text{
\ }\Delta =\nabla ^{2}
\end{equation}
where the kinetic terms involving the operation $\sqrt{-\Delta +m_{i}^{2}}$
are nonlocal operators and $\chi ({\bf r})=Y_{\ell ,m}(\theta ,\phi )R_{n,l}(%
{\bf r})$ denotes the Salpeter's wave function. For heavy interacting
particles, the kinetic energy operators in Eq. (11) can be approximated,
(cf. e.g., Jaczko and Durand [44], Ikhdair and Sever [45,46]), as

\begin{equation}
\mathrel{\mathop{\sum }\limits_{i=1,2}}%
\sqrt{-\Delta +m_{i}^{2}}=m_{1}+m_{2}-\frac{\Delta }{2\mu }-\frac{\Delta ^{2}%
}{8\eta ^{3}}-\;\cdots ,
\end{equation}
where $\mu =\frac{m_{1}m_{2}}{m_{1}+m_{2}}$ stands for the reduced mass and $%
\eta =\mu \left( \frac{m_{1}m_{2}}{m_{1}m_{2}-3\mu ^{2}}\right) ^{1/3}$\ is
an introduced useful mass parameter. This SS-type equation retains its
relativistic kinematics and is suitable for describing the spin-averaged
spectrum of two bound particles of masses $m_{1}$ and $m_{2}.$ The
Hamiltonian containing the relativistic corrections up to order $%
(v^{2}/c^{2})$ is called a generalized Breit-Fermi Hamiltonian (cf. e.g.,
Lucha {\it et al. }[47]). Hence, the SS equation can be further written in
the form ($\hbar =c=1)$ [45,46]
\begin{equation}
\left\{ -\frac{\Delta }{2\mu }-\frac{\Delta ^{2}}{8\eta ^{3}}+V(r)\right\}
R_{nl}({\bf r})=E_{nl}R_{nl}({\bf r}),
\end{equation}
where $E_{nl}=M_{nl}-m_{1}-m_{2}$ refers to the Salpeter binding energy with
$M_{nl}$ is the semirelativistic-bound-state masses.\footnote{%
This approximation is correct to $O(v^{2}/c^{2}).$ The $\Delta ^{2}$ term in
(13) should be properly treated as a perturbation by using trial
wavefunctions [48].} Moreover, it is worthwhile to point out here that, to
obtain a Schr\"{o}dinger-type equation, as could be seen fron Eq.(13), the
perturbation term can be treated by using the reduced Schr\"{o}dinger
equation [49]

\begin{equation}
p^{4}=4\mu ^{2}\left[ E_{nl}-V(r)\right] ^{2},
\end{equation}
with $p^{4}=\Delta ^{2},$ and consequently one would reduce Eq. (13) into a
Schr\"{o}dinger-type form [45,46]

\begin{equation}
\left\{ -\frac{\Delta }{2\mu }-\frac{\mu ^{2}}{2\eta ^{3}}\left[
E_{nl}^{2}+V^{2}(r)-2E_{nl}V(r)\right] +V(r)\right\} R_{nl}({\bf r}%
)=E_{nl}R_{nl}({\bf r}).
\end{equation}
In addition, the three-dimensional $(3D)$ space operator in the spherical
polar coordinates takes the form \
\begin{equation}
\Delta =\frac{\partial ^{2}}{\partial r^{2}}+\frac{2}{r}\frac{\partial }{%
\partial r}-\frac{L^{2}}{r^{2}},
\end{equation}
with $L^{2}=l\left( l+1\right) .$ Hence, after employing the following
transformation\ \ \
\begin{equation}
R_{nl}(r)=\frac{\psi _{nl}(r)}{r},
\end{equation}
we obtain [45]
\begin{equation}
\Delta =\frac{\partial ^{2}}{\partial r^{2}}-\frac{l\left( l+1\right) }{r^{2}%
},\text{ \ }\Delta ^{2}=\frac{\partial ^{4}}{\partial r^{4}}-\frac{2l\left(
l+1\right) }{r^{2}}\frac{\partial ^{2}}{\partial r^{2}}+\frac{4l\left(
l+1\right) }{r^{3}}\frac{\partial }{\partial r}+\frac{l\left( l+1\right) %
\left[ l^{2}+l-6\right] }{r^{4}}.
\end{equation}
Thus, using Eqs (17) and (18) and after a lengthy algebra but
straightforward, we can finally write Eq.(15) in the following $3D$ space

\begin{equation}
\left[ -\frac{\hbar ^{2}}{2\mu }\frac{d^{2}}{dr^{2}}+\frac{l\left(
l+1\right) \hbar ^{2}}{2\mu r^{2}}+W_{nl}(r)-\frac{W_{nl}(r)^{2}}{2%
\widetilde{m}}\right] \psi _{nl}(r)=0,
\end{equation}
where
\begin{equation}
W_{nl}(r)=V(r)-E_{nl},\text{ \ }\widetilde{m}=\eta ^{3}/\mu
^{2}=(m_{1}m_{2}\mu )/(m_{1}m_{2}-3\mu ^{2}).
\end{equation}

It is worthwhile to point out that the Schr\"{o}dinger-type equation (19),
is found to be same as the formula given by Durand and Durand [50]. The
perturbation term, $W_{nl}(r)^{2};$ that is, $(v^{2}/c^{2})$ term in Eqs.
(19) and (20) is significant only when it is small (i.e., $W_{nl}(r)/%
\widetilde{m}\ll 1)$. This condition is verified by the confining potentials
used to describe the present system except near the color$-$Coulomb
singularity at the origin, and for $r\rightarrow \infty $ (i.e., the
wavefunction vanishes at $0$ and $\infty $). However, it is always being
satisfied on the average as stated by Ref.[50].

In the present work, we shall study the SS equation with a family of
exponential potentials which is called generalized Hulth\'{e}n potential
[51], one of the important molecular potentials, in the $1D$ -vector form,

\begin{equation}
V_{q}(x)=-V_{0}\frac{e^{-\alpha x}}{1-qe^{-\alpha x}},\text{ \ }%
V_{0}=Ze^{2}\alpha ,\text{ \ }0\leq x\leq \infty
\end{equation}
with $\alpha $ denotes the screening (range) parameter, $V_{0}$ denotes the
coupling constant and $q$ is the deformation parameter which is used to
determine the shape of potential. We have to note that, for some specific $q$
values this potential reduces to the well-known types: such as for $q=0,$ to
the exponential potential, for $q=1$ to the standard Hulth\'{e}n potential,
and for $q=-1$ to the Woods-Saxon potential [41]. Further, near the origin,
it reduces into the shifted linear potential in the limit of very short
range (i.e., $\alpha \rightarrow 0)$ [31,32]
\begin{equation}
V_{q}(x)\approx \frac{V_{0}}{q-1}+\frac{V_{0}}{(q-1)^{2}}\alpha x+O(\alpha
^{2}x^{2}),
\end{equation}
with a constant shift, $V_{0}/(q-1)$ \ It also approximates to the screened
Coulomb effective potential for small $\alpha x$ (i.e., $\alpha x\rightarrow
0$) as [34]

\begin{equation}
V_{sc}^{eff}(x)\approx -\frac{e^{-\alpha x}}{x}+\frac{\ell (\ell +1)}{2x^{2}}%
.
\end{equation}
The complex form of the potential in (21) is said to be a ${\cal PT}${\rm \ -%
}Symmetric potential when (cf. Refs.[7,31,32,41])

\begin{equation}
\lbrack {\cal PT}\text{ ,}V(x)]=0,
\end{equation}
i.e., the ${\cal PT}{\rm -}$symmetry condition for the given potential $V(x)$
satisfies

\begin{equation}
\lbrack V(-x)]^{\ast }=V(x).
\end{equation}
To calculate the energy eigenvalues and the corresponding eigenfunctions,
the Hermitian real-valued Hulth\'{e}n potential form given by Eq.(21) is
substituted into the $1D$ ${\cal PT}{\rm -}$symmetrical Hermitian
Schr\"{o}dinger-type equation (19) for $\ell =0$ case $($i.e., $s$-wave
states$)$:

\begin{equation}
\frac{d^{2}\psi _{nq}(x)}{dx^{2}}+\frac{2\mu }{\hbar ^{2}}\left[ E_{n}+\frac{%
E_{n}^{2}}{2\widetilde{m}}+\frac{V_{0}e^{-\alpha x}}{(1-qe^{-\alpha x})}+%
\frac{V_{0}^{2}e^{-2\alpha x}}{2\widetilde{m}(1-qe^{-\alpha x})^{2}}+\frac{%
V_{0}E_{n}e^{-\alpha x}}{\widetilde{m}\left( 1-qe^{-\alpha x}\right) }\right]
\psi _{nq}(x)=0,
\end{equation}
where $\mu =m/2$ and $\widetilde{m}=2m$ for two identical interacting
particles. Now, introducing a convenient dimensionless transformation, s$%
(x)=e^{-\alpha x}$ , satisfying the arbitrary boundary conditions, $0\leq
x\leq \infty \rightarrow 1\leq s\leq 0,$ reduces Eq.(26) to the form:

\begin{equation}
\frac{d^{2}\psi _{nq}(s)}{ds^{2}}+\frac{1}{s}\frac{d\psi _{nq}(s)}{ds}+\frac{%
2\mu }{\hbar ^{2}\alpha ^{2}s^{2}}\left[ E_{n}+\frac{E_{n}^{2}}{2\widetilde{m%
}}+\frac{V_{0}s}{(1-qs)}+\frac{V_{0}^{2}s^{2}}{2\widetilde{m}(1-qs)^{2}}+%
\frac{V_{0}E_{n}s}{\widetilde{m}\left( 1-qs\right) }\right] \psi _{nq}(s)=0,
\end{equation}
with the dimensionless definitions given by
\[
-\epsilon ^{2}=\frac{2\mu }{\hbar ^{2}\alpha ^{2}}\left( E_{n}+\frac{%
E_{n}^{2}}{2\widetilde{m}}\right) \geq 0\text{ \ \ (}E_{n}\leq 0),\text{ \ \
}\epsilon _{1}=\frac{2\mu }{\hbar ^{2}\alpha ^{2}}V_{0}\text{ \ \ (}\epsilon
_{1}>0),\text{ \ }
\]
\begin{equation}
\epsilon _{2}^{2}=\frac{2\mu }{\hbar ^{2}\alpha ^{2}}\frac{V_{0}^{2}}{2%
\widetilde{m}}\text{ \ (}\epsilon _{2}^{2}>0),\text{ \ }\epsilon _{3}=\frac{%
2\mu }{\hbar ^{2}\alpha ^{2}}\frac{V_{0}E_{n}}{\widetilde{m}}\text{ \ (\ }%
\epsilon _{3}>0)
\end{equation}
and finally one can arrive at the simple hypergeometric equation given by

\begin{equation}
\psi {}_{nq}^{\prime \prime }(s)+\frac{1-qs}{s(1-qs)}\psi {}_{nq}^{\prime
}(s)+\frac{\left[ s^{2}(\epsilon _{2}^{2}-q^{2}\epsilon ^{2}-q\epsilon
_{3}-q\epsilon _{1})+s\left( \epsilon _{1}+\epsilon _{3}+2q\epsilon
^{2}\right) -\epsilon ^{2}\right] }{\left[ s\left( 1-qs\right) \right] ^{2}}%
\psi _{nq}(s)=0.
\end{equation}
Hence, comparing the last equation with the generalized hypergeometric type,
Eq.(1), we obtain the associated polynomials as

\begin{equation}
\widetilde{\tau }(s)=1-qs,\text{ \ \ \ }\sigma (s)=s(1-qs),\text{ \ \ }%
\widetilde{\sigma }(s)=s^{2}(\epsilon _{2}^{2}-q^{2}\epsilon ^{2}-q\epsilon
_{3}-q\epsilon _{1})+s\left( \epsilon _{1}+\epsilon _{3}+2q\epsilon
^{2}\right) -\epsilon ^{2}.
\end{equation}
When these polynomials are substituted into Eq.(7), with $\sigma ^{\prime
}(s)=1-2qs,$ we obtain

\begin{equation}
\pi (s)=-\frac{qs}{2}\pm \frac{1}{2}\sqrt{s^{2}\left( q^{2}+4(q^{2}\epsilon
^{2}+q\epsilon _{1}+q\epsilon _{3}-\epsilon _{2}^{2})-4qk\right) +4s\left(
k-\epsilon _{1}-\epsilon _{3}-2q\epsilon ^{2}\right) +4\epsilon ^{2}}.
\end{equation}
Further, the discriminant of the upper expression under the square root has
to be set equal to zero. Therefore, it becomes

\begin{equation}
\Delta =\left[ k-\epsilon _{1}-\epsilon _{3}-2q\epsilon ^{2}\right]
^{2}-\epsilon ^{2}\left[ q^{2}+4(q^{2}\epsilon ^{2}+q\epsilon _{1}+q\epsilon
_{3}-\epsilon _{2}^{2})-4qk\right] =0.
\end{equation}
Solving Eq.(32) for the constant $k,$ we obtain the double roots as $%
k_{+,-}=\epsilon _{1}+\epsilon _{3}\pm b\epsilon ,$ where $b=\sqrt{%
q^{2}-4\epsilon _{2}^{2}}.$ Thus, substituting these values for each $k$
into Eq.(31), we obtain

\begin{equation}
\pi (s)=-\frac{qs}{2}\pm \frac{1}{2}\left\{
\begin{array}{c}
\left( 2q\epsilon -b\right) s-2\epsilon ;\text{ \ \ \ for \ \ }%
k_{+}=\epsilon _{1}+\epsilon _{3}+b\epsilon , \\
\left( 2q\epsilon +b\right) s-2\epsilon ;\text{ \ \ for \ \ }k_{-}=\epsilon
_{1}+\epsilon _{3}-b\epsilon .
\end{array}
\right.
\end{equation}
Hence, making the following choice for the polynomial $\pi (s)$ as

\begin{equation}
\pi (s)=-\frac{qs}{2}-\frac{1}{2}\left[ \left( 2q\epsilon +b\right)
s-2\epsilon \right] ,
\end{equation}
for $k_{-}=\epsilon _{1}+\epsilon _{3}-b\epsilon ,$ giving the function:

\begin{equation}
\tau \text{(s)}=-q(2+2\epsilon +b/q)s+(1+2\epsilon ),
\end{equation}
which has a \ negative derivative of the form $\tau
{\acute{}}%
(s)=-q(2+2\epsilon +b/q).$ Thus, from Eqs.(8)-(9) and Eqs.(34)-(35), we find

\begin{equation}
\lambda =-\frac{q}{2}(1+2\epsilon )\left( 1+b/q\right) +(\epsilon
_{1}+\epsilon _{3}),
\end{equation}
and

\begin{equation}
\lambda _{n}=\left( 1+n+2\epsilon +b/q\right) nq.
\end{equation}
Therefore, after setting $\lambda _{n}=\lambda $ and solving for $\epsilon ,$
in $\ \hbar =1$ units, we find the Salpeter exact binding energy spectra as

\begin{equation}
E_{nq}=\left( \frac{V_{0}}{2q}-\widetilde{m}\right) \left\{ 1\pm \sqrt{1-%
\frac{2\widetilde{m}a}{q}\frac{\left[ \left( \frac{V_{0}}{a}\right)
^{2}-\left( \frac{V_{0}}{2a}\right) D+\frac{1}{16}D^{2}\right] }{\left(
\frac{V_{0}}{2q}-\widetilde{m}\right) ^{2}D}}\right\} ,\text{ \ }a=\frac{%
\hbar ^{2}\alpha ^{2}}{2\mu },\text{ }
\end{equation}
where

\begin{equation}
D=\left( C^{2}+4\epsilon _{2}^{2}\right) /q=q+q(2n+1)^{2}+2(2n+1)b,\text{ \ }%
C=b+q(2n+1)
\end{equation}
where $b=\sqrt{q^{2}-\frac{V_{0}^{2}}{\alpha ^{2}}}$ for equal mass case.
For convenience, when $m_{1}=m_{2},$ the upper expression (38) can be
further rearranged as

\begin{equation}
E_{nq}=\left( \frac{V_{0}}{2q}-2m\right) \left\{ 1\pm \sqrt{1-\frac{\left(
2mV_{0}\right) ^{2}}{\xi }\frac{\left[ 1-\left( \frac{\xi }{2mqV_{0}}\right)
+\frac{1}{4}\left( \frac{\xi }{2mqV_{0}}\right) ^{2}\right] }{\left( \frac{%
V_{0}}{2q}-2m\right) ^{2}}}\right\} ,\text{ \ \ }0\leq n<\infty ,\text{ \ \ }
\end{equation}
where

\[
\xi =q\alpha \left[ q\alpha +q\alpha (2n+1)^{2}+2(2n+1)\sqrt{q^{2}\alpha
^{2}-V_{0}^{2}}\right] =\kappa ^{2}+V_{0}^{2},
\]
\begin{equation}
\text{\ }\kappa =\sqrt{\alpha ^{2}q^{2}-V_{0}^{2}}+q\alpha (2n+1),\text{ \ }%
q^{2}\geq \left( \frac{V_{0}}{\alpha }\right) ^{2}.
\end{equation}
Let us now find the corresponding wavefunctions. Applying the ${\rm NU}$
method, the hypergeometric function $y_{n}(s)$ is the polynomial solution of
hypergeometric-type equation (3) and described with the weight function
[42]. By substituting $\pi (s)$ and $\sigma (s)$ in Eq.(6) and then solving
the first-order differential equation, we find

\begin{equation}
\phi (s)=s^{\epsilon }(1-qs)^{\frac{(b+q)}{2q}}.\text{ }
\end{equation}
It is easy to find the other part of the wave function from the definition
of the weight function

\begin{equation}
\rho (s)=s^{2\epsilon }(1-qs)^{b/q},
\end{equation}
which then substituted into the Rodrigues relation resulting in

\begin{equation}
y_{nq}(s)=D_{nq}s^{-2\epsilon }(1-qs)^{-b/q}\frac{d^{n}}{ds^{n}}\left[
s^{n+2\epsilon }\left( 1-qs\right) ^{n+b/q}\right] ,
\end{equation}
where $D_{nq}$ is a normalizing constant$.$ In the limit $q\rightarrow 1,$
the polynomial solutions of $\ y_{n}(s)$ are expressed in terms of Jacobi
Polynomials, which is one of the classical orthogonal polynomials, with
weight function (43) in the closed interval $\left[ 0,1\right] ,$ giving $%
y_{n,1}(s)\simeq P_{n}^{(2\epsilon ,b)}(1-2s)$ [52]. The radial wave
function $\psi _{nq}(s)$ is obtained from the Jacobi polynomials in Eq.(44)
and $\phi (s)$ in Eq.(42) for the ${\rm s}$-wave functions could be
determined as

\begin{equation}
\psi _{nq}(s)=N_{nq}s^{-\epsilon }(1-qs)^{\frac{q-b}{2q}}\frac{d^{n}}{ds^{n}}%
\left[ s^{n+2\epsilon }\left( 1-qs\right) ^{n+b/q}\right] =N_{nq}s^{\epsilon
}(1-qs)^{\frac{(b+q)}{2q}}P_{n}^{(2\epsilon ,b/q)}(1-2qs),
\end{equation}
with $s=e^{-\alpha x}$ and $N_{nq}$ is a new normalization constant
determine by

\begin{equation}
1=\int\limits_{1}^{0}\left| \psi _{nq}(s)\right|
^{2}ds=N_{nq}^{2}\int\limits_{1}^{0}s^{2\epsilon }(1-qs)^{\frac{(q+b)}{q}}%
\left[ P_{n}^{(2\epsilon ,b/q)}(1-2qs)\right] ^{2}ds,
\end{equation}
We now make use of the fact that the Jacobi polynomials can be explicitly
written in two different ways [52]:

\begin{equation}
P_{n}^{(\rho ,\nu )}(z)=2^{-n}\sum\limits_{p=0}^{n}(-1)^{n-p}%
{n+\rho  \choose p}%
{n+\nu  \choose n-p}%
\left( 1-z\right) ^{n-p}\left( 1+z\right) ^{p},
\end{equation}

\begin{equation}
P_{n}^{(\rho ,\nu )}(z)=\frac{\Gamma (n+\rho +1)}{n!\Gamma (n+\rho +\nu +1)}%
\sum\limits_{r=0}^{n}%
{n \choose r}%
\frac{\Gamma (n+\rho +\nu +r+1)}{\Gamma (r+\rho +1)}\left( \frac{z-1}{2}%
\right) ^{r},
\end{equation}
where $%
{n \choose r}%
=\frac{n!}{r!(n-r)!}=\frac{\Gamma (n+1)}{\Gamma (r+1)\Gamma (n-r+1)}.$ Using
Eqs.(47)-(48), we obtain the explicit expressions for $P_{n}^{(2\epsilon
,b/q)}(1-2qs)$

\[
P_{n}^{(2\epsilon ,b/q)}(1-2qs)=(-1)^{n}\Gamma (n+2\epsilon +1)\Gamma (n+%
\frac{b}{q}+1)
\]

\begin{equation}
\times \sum\limits_{p=0}^{n}\frac{(-1)^{p}q^{n-p}}{p!(n-p)!\Gamma (p+\frac{b%
}{q}+1)\Gamma (n+2\epsilon -p+1)}s^{n-p}(1-qs)^{p},
\end{equation}

\begin{equation}
P_{n}^{(2\epsilon ,b/q)}(1-2qs)=\frac{\Gamma (n+2\epsilon +1)}{\Gamma
(n+2\epsilon +\frac{b}{q}+1)}\sum\limits_{r=0}^{n}\frac{(-1)^{r}q^{r}\Gamma
(n+2\epsilon +\frac{b}{q}+r+1)}{r!(n-r)!\Gamma (2\epsilon +r+1)}s^{r}.
\end{equation}
Therefore, substituting Eqs.(49) and (50) into Eq.(46), it gives
\[
1=N_{nq}^{2}(-1)^{n+1}\frac{\Gamma (n+\frac{b}{q}+1)\Gamma (n+2\epsilon
+1)^{2}}{\Gamma (n+2\epsilon +\frac{b}{q}+1)}\left\{ \sum\limits_{p=0}^{n}%
\frac{(-1)^{p}q^{n-p}}{p!(n-p)!\Gamma (p+\frac{b}{q}+1)\Gamma (n+2\epsilon
-p+1)}\right\}
\]

\begin{equation}
\times \left\{ \sum\limits_{r=0}^{n}\frac{(-1)^{r}q^{r}\Gamma (n+2\epsilon +%
\frac{b}{q}+r+1)}{r!(n-r)!\Gamma (2\epsilon +r+1)}\right\} I_{nq}(p,r),
\end{equation}
where

\begin{equation}
I_{nq}(p,r)=\int\limits_{0}^{1}s^{n+2\epsilon +r-p}(1-qs)^{p+\frac{b}{q}%
+1}ds.
\end{equation}
Using the following integral representation of the hypergeometric function
[53]

\[
\int\limits_{0}^{1}s^{\alpha _{0}-1}(1-s)^{\gamma _{0}-\alpha
_{0}-1}(1-qs)^{-\beta _{0}}ds=_{2}F_{1}(\alpha _{0},\beta _{0}:\gamma _{0};q)%
\frac{\Gamma (\alpha _{0})\Gamma (\gamma _{0}-\alpha _{0})}{\Gamma (\gamma
_{0})},
\]

\begin{equation}
\lbrack
\mathop{\rm Re}%
(\gamma _{0})>%
\mathop{\rm Re}%
(\alpha _{0})>0,\text{ \ }\left| \arg (1-q)\right| <\pi ]
\end{equation}
which gives

\begin{equation}
_{2}F_{1}(\alpha _{0},\beta _{0}:\alpha _{0}+1;q)/\alpha
_{0}=\int\limits_{0}^{1}s^{\alpha _{0}-1}(1-qs)^{-\beta _{0}}ds.
\end{equation}

The hypergeometric function $_{2}F_{1}(\alpha _{0},\beta _{0}:\gamma _{0};1)$
reduces into

\[
_{2}F_{1}(\alpha _{0},\beta _{0}:\gamma _{0};1)=\frac{\Gamma (\gamma
_{0})\Gamma (\gamma _{0}-\alpha _{0}-\beta _{0})}{\Gamma (\gamma _{0}-\alpha
_{0})\Gamma (\gamma _{0}-\beta _{0})},
\]
\begin{equation}
\lbrack
\mathop{\rm Re}%
(\gamma _{0}-\alpha _{0}-\beta _{0})>0,\text{ }%
\mathop{\rm Re}%
(\gamma _{0})>%
\mathop{\rm Re}%
(\beta _{0})>0],
\end{equation}
for $q=1.$ Setting $\alpha _{0}=n+2\epsilon +r-p+1,$ $\beta _{0}=-p-\frac{b}{%
q}-1,$ and $\gamma _{0}=\alpha _{0}+1,$ one gets

\begin{equation}
I_{nq}(p,r)=\frac{_{2}F_{1}(\alpha _{0},\beta _{0}:\gamma _{0};q)}{\alpha
_{0}}=\frac{(n+2\epsilon +r-p+1)!(p+\frac{b}{q}+1)!}{(n+2\epsilon
+r-p+1)(n+2\epsilon +r+\frac{b}{q}+2)!}.
\end{equation}

\subsection{Real potentials}

Firstly, we consider the real case in Eq.(21), i.e., all parameters $%
(V_{0},q,\alpha )$ are real:

(i) For any given $\alpha $ the spectrum consists of real eigenstate spectra
$E_{n}(V_{0},q,\alpha )$ depending on $q.$ The sign of $V_{0}$ does not
affect the bound states. It is clear that while $V_{0}\rightarrow 0,$ $%
E_{n}=-2m\left[ 1\pm \sqrt{1-\left( \frac{(n+1)\alpha }{2m}\right) ^{2}}%
\right] $ which is for the ground state (i.e., $n=0)$ tend to the value $%
E_{0}\approx -3.73m$ and for the first excited state (i.e., $n=1)$ takes the
value $E_{1}=-2m,$ where we have used $\lambda _{c}=\frac{\hbar }{mc}=\frac{1%
}{m}=\frac{1}{\alpha \text{ }}$ which is the compton wavelength of the
Salpeter particles.

(ii) There exist bound states (real solution) in case if the condition $%
\frac{V_{0}^{2}}{\alpha ^{2}}\leq q^{2}$ is achieved, otherwise there are no
bound-states.

(iii) There exist bound states in case if the condition $\left(
2mV_{0}\right) ^{2}\left[ 1-\frac{\xi }{2mqV_{0}}+\frac{1}{4}\left( \frac{%
\xi }{2mqV_{0}}\right) ^{2}\right] \leq \left( \frac{V_{0}}{2q}-2m\right)
^{2}\xi ,$ with $\xi =\kappa ^{2}+V_{0}^{2}$ is achieved, otherwise there
are no bound-states.

Moreover, this condition which gives the critical coupling value turns to be

\begin{equation}
n\leq \frac{1}{2q\alpha }\left( \sqrt{\chi ^{2}-V_{0}^{2}}-\sqrt{q^{2}\alpha
^{2}-V_{0}^{2}}\right) -\frac{1}{2},
\end{equation}
where

\begin{equation}
\chi ^{2}=2q^{2}\left\{ \left[ \frac{2mV_{0}}{q}+\left( \frac{V_{0}}{2q}%
-2m\right) ^{2}\right] \pm \left( \frac{V_{0}}{2q}-2m\right) \sqrt{\left(
\frac{V_{0}}{2q}-2m\right) ^{2}+\frac{4mV_{0}}{q}}\right\}
\end{equation}
i.e., there are only finitely many eigenstates. In order that at least one
level might exist, its necessary that the inequality

\begin{equation}
q\alpha +\sqrt{q^{2}\alpha ^{2}-V_{0}^{2}}\leq \sqrt{\chi ^{2}-V_{0}^{2}},
\end{equation}
is fulfilled

For a more specific case $q=-1,$ Eq.(21) is reduced into the shifted
Woods-Saxon (WS) potential

\begin{equation}
V(x)=-V_{0}+\frac{V_{0}}{1+e^{-\alpha x}},
\end{equation}
with energy spectrum given by

\begin{equation}
E_{n}(V_{0},\alpha )=-\left( \frac{V_{0}}{2q}+2m\right) \left\{ 1\pm \sqrt{1-%
\frac{\left( 2mV_{0}\right) ^{2}}{\widetilde{\xi }}\frac{\left[ 1+\left(
\frac{\widetilde{\xi }}{2mV_{0}}\right) +\frac{1}{4}\left( \frac{\widetilde{%
\xi }}{2mV_{0}}\right) ^{2}\right] }{\left( \frac{V_{0}}{2}+2m\right) ^{2}}}%
\right\} ,\text{ \ \ }0\leq n<\infty ,\text{ \ \ }
\end{equation}
where

\begin{equation}
\widetilde{\xi }=\alpha \left[ \alpha +\alpha (2n+1)^{2}-2(2n+1)\sqrt{\alpha
^{2}-V_{0}^{2}}\right] ,\text{ \ }\alpha ^{2}\geq V_{0}^{2}.
\end{equation}
In this case, for any given $\alpha $, all the eigenstates $E_{n}\leq 0.$

(iv) For $q=0,$ the potential (21) reduces into the exponential form

\begin{equation}
V(x)=-V_{0}e^{-\alpha x},
\end{equation}
the energy expression (40) does not give an explicit form, i.e., the NU
method fails to give energy expression to this type of exponential
potential. It is noted that for this potential there is no explicit form of
energy expression of bound states for Schr\"{o}dinger [16], KG [31,38] and
also Dirac [36] equations.

For $q=0,$ the generalized equation of hypergeometric type which is given by
Eq.(29) becomes

\begin{equation}
\psi {}_{n}^{\prime \prime }(s)+\frac{1}{s}\psi {}_{n}^{\prime }(s)+\frac{%
\left[ s^{2}\epsilon _{2}^{2}+s\left( \epsilon _{1}+\epsilon _{3}\right)
-\epsilon ^{2}\right] }{s^{2}}\psi _{n}(s)=0,
\end{equation}
with

\begin{equation}
\widetilde{\tau }(s)=1,\text{ \ \ \ }\sigma (s)=s,\text{ \ \ }\widetilde{%
\sigma }(s)=s^{2}\epsilon _{2}^{2}+s(\epsilon _{1}+\epsilon _{3})-\epsilon
^{2},
\end{equation}
and the corresponding $\pi (s)$ is determined as

\begin{equation}
\pi (s)=\pm \left\{
\begin{array}{c}
i\epsilon _{2}s+\epsilon ;\text{ \ \ \ for \ \ }k_{+}=\epsilon _{1}+\epsilon
_{3}+2i\epsilon _{2}\epsilon , \\
i\epsilon _{2}s-\epsilon ;\text{ \ \ for \ \ }k_{-}=\epsilon _{1}+\epsilon
_{3}-2i\epsilon _{2}\epsilon ,
\end{array}
\right.
\end{equation}
where $i=\sqrt{-1}$ and $\epsilon _{2}=\frac{V_{0}}{2\alpha }.$ Following a
procedure similar to the previous case, when $\pi (s)=-i\epsilon
_{2}s+\epsilon $ is chosen for $k=\epsilon _{1}+\epsilon _{3}-2i\epsilon
_{2}\epsilon ,$ then

\begin{equation}
\tau (s)=-2i\epsilon _{2}s+(1+2\epsilon ),
\end{equation}
and

\begin{equation}
\lambda =\epsilon _{1}+\epsilon _{3}-i\epsilon _{2}-2i\epsilon _{2}\epsilon ,%
\text{ }\lambda _{n}=2ni\epsilon _{2},\text{\ }\phi _{n}(s)=s^{\epsilon
}e^{-i\epsilon _{2}s},
\end{equation}
could be obtained. Substituting $\sigma (s)$ and $\tau (s),$ together with $%
\lambda $ into Eq.(3) gives

\begin{equation}
sy_{n}^{\prime \prime }(s)+\left[ 1+2\epsilon -2i\epsilon _{2}s\right]
y_{n}^{\prime }(s)-\left[ i\epsilon _{2}+2i\epsilon _{2}\epsilon -(\epsilon
_{1}+\epsilon _{3})\right] y_{n}(s)=0.
\end{equation}
The last equation can also be reduced to standard Whittaker differential
equation [54]. Thus, the solutions vanishing at infinity and it can be
written in terms of confluent hypergeometric function as follows:

\begin{equation}
y_{n}(s)=_{1}F_{1}\left( \frac{1}{2}+\epsilon +i\frac{(\epsilon
_{1}+\epsilon _{3})}{2\epsilon _{2}};1+2\epsilon ;2i\epsilon _{2}s\right) .
\end{equation}
So, the acceptable solution for the upper component is found to be

\begin{equation}
\psi _{n}(s)=\phi (s)y_{n}(s)=A_{1}F_{1}\left( \frac{1}{2}+\epsilon +i\frac{%
(\epsilon _{1}+\epsilon _{3})}{2\epsilon _{2}};1+2\epsilon ;2i\epsilon
_{2}s\right) s^{\epsilon }e_{1}^{-i\epsilon _{2}s}.
\end{equation}
On the other hand, to find the an expression for the exact energy spectrum,
when $\lambda _{n}=\lambda ,$ in $\ \hbar =1$ units, we obtain

\begin{equation}
E_{n}(\alpha )=i\frac{2n+1}{2}\alpha -2m-i\frac{2m^{2}}{(2n+1)\alpha },\text{
\ \ }0\leq n<\infty .\text{ \ \ }
\end{equation}
>From the last equation we conclude that if and only if $i\alpha $ is real
then $E_{n}$ is a real. Hence, in order to apply NU method to this type of
exponential potential $\epsilon _{2}$ should be complex (imaginary). This
leads to result that either $V_{0}$ or $\alpha $ must be imaginary.
Therefore, for purely imaginary $\alpha ,$ i.e., $\alpha \rightarrow i\alpha
_{I},$ it reads:

\begin{equation}
E_{n}(\alpha )=-m\left[ 2+\frac{(2n+1)}{2m}\alpha _{I}+\frac{2m}{%
(2n+1)\alpha _{I}}\right] ,\text{ \ \ }0\leq n<\infty .
\end{equation}
In the nonrelativistic limit, there is no available bound-state energy
solution for the exponential potentials [16].

\subsection{Complex potentials}

Let us consider the case where at least one of the potential parameters be
complex:

(I) If $\alpha $ is a complex parameter ($\alpha \rightarrow i\alpha $), the
potential (21) becomes

\begin{equation}
V_{q}(x)=\frac{V_{0}}{q^{2}-2q\cos (\alpha x)+1}\left[ q-\cos (\alpha
x)+i\sin (\alpha x)\right] =V_{q}^{\ast }(-x),
\end{equation}
which is a ${\cal PT}{\rm -}$symmetric but non-Hermitian. It has real
spectrum given by

\[
E_{n}(V_{0},i\alpha ,q)=\left( \frac{V_{0}}{2q}-2m\right) \left\{ 1\pm \sqrt{%
1+\frac{\left( 2mV_{0}\right) ^{2}}{\varsigma }\frac{\left[ 1+\left( \frac{%
\varsigma }{2mqV_{0}}\right) +\frac{1}{4}\left( \frac{\varsigma }{2mqV_{0}}%
\right) ^{2}\right] }{\left( \frac{V_{0}}{2q}-2m\right) ^{2}}}\right\} ,
\]
with

\begin{equation}
\varsigma =q^{2}\alpha ^{2}+q^{2}\alpha ^{2}(2n+1)^{2}+2q\alpha (2n+1)\sqrt{%
q^{2}\alpha ^{2}+V_{0}^{2}},
\end{equation}
if and only if $\ -\left( 2mV_{0}\right) ^{2}\left[ 1+\left( \frac{\varsigma
}{2mV_{0}q}\right) +\frac{1}{4}\left( \frac{\varsigma }{2mqV_{0}}\right) ^{2}%
\right] \leq \varsigma \left( \frac{V_{0}}{2q}-2m\right) ^{2}.$ The
corresponding radial wave function $\psi _{nq}(s)$ for the ${\rm s}$-wave
could be determined as

\begin{equation}
\psi _{nq}(s)=N_{nq}s^{i\epsilon }(1-qs)^{\frac{(c+q)}{2q}%
}P_{n}^{(2i\epsilon ,c/q)}(1-2qs),
\end{equation}
where $c=\sqrt{q^{2}+\frac{V_{0}^{2}}{\alpha ^{2}}},$ and $s=e^{-i\alpha x}.$

The norm of the wavefunction of such a non-Hermitian quantum mechanical
system is redefined as [2,55]

\begin{equation}
\int\limits_{0}^{\infty }\psi _{nq}^{\ast }(-s)\psi _{nq}(s)ds=\nu ,\text{ \
\ \ }\nu =\pm 1.
\end{equation}
$\nu =1$ stands for ${\cal PT}{\rm -}$symmetric phase whereas $\nu =-1$
stands for ${\cal PT}{\rm -}$antisymmetric phase. Therefore making the
necessary parameter replacements in Eqs.(51)-(52), we can obtain the
normalization constant for the complex ${\cal PT}{\rm -}$symmetric
generalized Hulth\'{e}n potential given by Eq.(74).

For the sake of comparing the relativistic and non-relativistic binding
energies, we need to solve the $1D$ \ Schr\"{o}dinger equation for the
complex generalized Hulth\'{e}n potential. We set the convenient
transformation $s(x)=e^{-i\alpha x}$ , $0\leq x\leq \infty \rightarrow 1\leq
s\leq 0,$ to obtain

\begin{equation}
\psi {}_{nq}^{\prime \prime }(s)+\frac{1-qs}{(s-qs^{2})}\psi {}_{nq}^{\prime
}(s)+\frac{\left[ s^{2}(q\widetilde{\epsilon }_{1}-q^{2}\widetilde{\epsilon }%
^{2}\text{\ })+s(2q\widetilde{\epsilon }^{2}-\widetilde{\epsilon }_{1})-%
\widetilde{\epsilon }^{2}\right] }{\left( s-qs^{2}\right) ^{2}}\psi
_{nq}(s)=0,
\end{equation}
for which
\[
\widetilde{\tau }(s)=1-qs,\text{ \ \ \ }\sigma (s)=s-qs^{2},\text{ \ \ }%
\widetilde{\sigma }(s)=s^{2}(q\widetilde{\epsilon }_{1}-q^{2}\widetilde{%
\epsilon }^{2}\text{\ })+s(2q\widetilde{\epsilon }^{2}-\widetilde{\epsilon }%
_{1})-\widetilde{\epsilon }^{2},
\]

\begin{equation}
\widetilde{\epsilon }^{2}=\frac{2\mu }{\hbar ^{2}\alpha ^{2}}E_{n},\text{ \ }%
\widetilde{\epsilon }_{1}=\frac{2\mu }{\hbar ^{2}\alpha ^{2}}V_{0}\text{ \ \
(}\widetilde{\epsilon }_{1}>0).
\end{equation}
Moreover, it could be obtained

\begin{equation}
\tau (s)=-q(3+2\widetilde{\epsilon })s+(1+2\widetilde{\epsilon }),
\end{equation}
if $\pi (s)=-q(1+\widetilde{\epsilon })s+\widetilde{\epsilon }$ is chosen
for $k_{-}=-q\widetilde{\epsilon }-\widetilde{\epsilon }_{1}.$ Finally, the
bound-state spectra in the non-relativistic limit could be found as

\begin{equation}
E_{n}(V_{0},q,i\alpha )=\frac{\hbar ^{2}}{8\mu q^{2}\alpha ^{2}}\left[ \frac{%
2\mu V_{0}/\hbar ^{2}+q\alpha ^{2}(n+1)^{2}}{(n+1)}\right] ^{2}>0,\text{ \ \
}0\leq n<\infty .\text{ \ \ }
\end{equation}
On the other hand, for the sake of comparison, the non-relativistic limit
for the real potential (21) can be found directly from the last equation as

\begin{equation}
E_{n}(V_{0},q,\alpha )=-\frac{\hbar ^{2}\alpha ^{2}}{8\mu }\left[ (n+1)-%
\frac{\beta }{(n+1)}\right] ^{2},\text{ \ \ }0\leq n<\infty .\text{ \ \ }
\end{equation}
where $\beta =\frac{2\mu V_{0}}{q\alpha ^{2}\hbar ^{2}}.$ The radial wave
function in the current case becomes

\begin{equation}
\psi _{nq}(s)=N_{nq}s^{\widetilde{\epsilon }}(1-qs)P_{n}^{(2\widetilde{%
\epsilon },1)}(1-2qs),
\end{equation}
with $s=e^{-i\alpha x}$ and $N_{nq}$ is a new normalization constant
determined by

\[
1=N_{nq}^{2}(-1)^{n+1}\frac{(n+1)!\Gamma (n+2\widetilde{\epsilon }+1)^{2}}{%
\Gamma (n+2\widetilde{\epsilon }+2)}\left\{ \sum\limits_{p=0}^{n}\frac{%
(-1)^{p}q^{n-p}}{p!(n-p)!(p+1)!\Gamma (n+2\widetilde{\epsilon }-p+1)}%
\right\}
\]

\begin{equation}
\times \left\{ \sum\limits_{r=0}^{n}\frac{(-1)^{r}q^{r}\Gamma (n+2\widetilde{%
\epsilon }+r+2)}{r!(n-r)!\Gamma (2\widetilde{\epsilon }+r+1)}\right\}
\int\limits_{0}^{1}s^{n+2\widetilde{\epsilon }+r-p}(1-qs)^{p+2}ds,
\end{equation}
where

\begin{equation}
\int\limits_{0}^{1}s^{n+2\widetilde{\epsilon }%
+r-p}(1-qs)^{p+2}ds=_{2}F_{1}(n+2\widetilde{\epsilon }+r-p+1,-p-2:n+2%
\widetilde{\epsilon }+r-p+2;q)B(n+2\widetilde{\epsilon }+r-p+1,1),
\end{equation}
(II) Let two parameters; namely, $V_{0}$ and $q$ be complex parameters
(i.e., $V_{0}\rightarrow iV_{0},$ $q\rightarrow iq$), then the potential
transforms to the form

\begin{equation}
V_{q}(x)=V_{0}\frac{\left[ 2\cosh ^{2}(\alpha x)-\sinh (2\alpha x)-1\right]
-i\left[ \cosh (\alpha x)-\sinh (\alpha x)\right] }{1+q^{2}\left[ 2\cosh
^{2}(\alpha x)-\sinh (2\alpha x)-1\right] },
\end{equation}
which is a ${\cal PT}{\rm -}$symmetric but non-Hermitian if $\left(
2mV_{0}\right) ^{2}\left[ 1-\left( \frac{\xi }{2mV_{0}q}\right) +\frac{1}{4}%
\left( \frac{\xi }{2mqV_{0}}\right) ^{2}\right] \leq \xi \left( \frac{V_{0}}{%
2q}-2m\right) ^{2},$ it may possesss real spectra as

\[
E_{n}(iV_{0},\alpha ,iq)=\left( \frac{V_{0}}{2q}-2m\right) \left\{ 1\pm
\sqrt{1-\frac{\left( 2mV_{0}\right) ^{2}}{\xi }\frac{\left[ 1-\left( \frac{%
\xi }{2mqV_{0}}\right) +\frac{1}{4}\left( \frac{\xi }{2mqV_{0}}\right) ^{2}%
\right] }{\left( \frac{V_{0}}{2q}-2m\right) ^{2}}}\right\} ,
\]
where $\xi $ is defined by Eq.(41). On the other hand, the corresponding
radial wave functions $\psi _{nq}(s)$ for the ${\rm s}$-wave could be
determined as

\begin{equation}
\psi _{nq}(s)=N_{nq}s^{\epsilon }(1-iqs)^{\frac{(d+q)}{2q}}P_{n}^{(2\epsilon
,d/q)}(1-2iqs),
\end{equation}
with $s=e^{-\alpha x}$ and $d=\sqrt{q^{2}-\frac{V_{0}^{2}}{\alpha ^{2}}.}$
The integral $I_{nq}(p,r)=\int\limits_{0}^{1}s^{n+2\epsilon +r-p}(1-iqs)^{p+%
\frac{d}{q}+1}ds$ is given by

\begin{equation}
I_{nq}(p,r)=_{2}F_{1}(n+2\epsilon +r-p+1,-p-\frac{d}{q}-1:n+2\epsilon
+r-p+2;iq)B(n+2\epsilon +r-p+1,1).
\end{equation}
(III) When all the parameters $V_{0},$ $\alpha $ and $q$ are complex
parameters (i.e., $V_{0}\rightarrow iV_{0},$ $\alpha \rightarrow i\alpha ,$ $%
q\rightarrow iq$), we obtain

\begin{equation}
V_{q}(x)=\frac{V_{0}}{q^{2}-2q\sin (\alpha x)+1}\left[ q-\sin (\alpha
x)-i\cos (\alpha x)\right] =V_{q}^{\ast }(\frac{\pi }{2}-x).
\end{equation}
This potential is a pseudo-Hermitian potential [27,56] having a $\pi /2$
phase difference with respect to the potential in case (I), it is also a $%
{\cal PT}{\rm -}$symmetric, $\eta =P$-pseudo-Hermitian (i.e., $%
PTV_{q}(x)(PT)^{-1}=V_{q}(x),$ with $P=\eta :x\rightarrow \frac{\pi }{%
2\alpha }-x$ and $T:i\rightarrow -i)$ but non-Hermitian having real spectrum
given by

\begin{equation}
E_{n}(iV_{0},i\alpha ,iq)=\left( \frac{V_{0}}{2q}-2m\right) \left\{ 1\pm
\sqrt{1+\frac{\left( 2mV_{0}\right) ^{2}}{\widetilde{\varsigma }}\frac{\left[
1+\left( \frac{\widetilde{\varsigma }}{2mqV_{0}}\right) +\frac{1}{4}\left(
\frac{\widetilde{\varsigma }}{2mqV_{0}}\right) ^{2}\right] }{\left( \frac{%
V_{0}}{2q}-2m\right) ^{2}}}\right\} ,
\end{equation}
where

\begin{equation}
\widetilde{\varsigma }=q^{2}\alpha ^{2}+q^{2}\alpha ^{2}(2n+1)^{2}-2q\alpha
(2n+1)\sqrt{q^{2}\alpha ^{2}+V_{0}^{2}}.
\end{equation}
On the other hand, the corresponding radial wave functions $\psi _{nq}(s)$
for the ${\rm s}$-wave could be determined as

\begin{equation}
\psi _{nq}(s)=N_{nq}s^{i\epsilon }(1-iqs)^{\frac{(c+q)}{2q}%
}P_{n}^{(2i\epsilon ,\widetilde{c}/q)}(1-2iqs),
\end{equation}
with $s=e^{-i\alpha x}$ and $c$ is defined after Eq.(76)$.$ The integral $%
I_{nq}(p,r)=\int\limits_{0}^{1}s^{n+2i\epsilon +r-p}(1-iqs)^{p+\frac{c}{q}%
+1}ds$ is given by

\begin{equation}
I_{nq}(p,r)=_{2}F_{1}(n+2i\epsilon +r-p+1,-p-\frac{c}{q}-1:n+2i\epsilon
+r-p+2;iq)B(n+2i\epsilon +r-p+1,1).
\end{equation}

\section{Results And Conclusions}

\label{RAC}We have seen that the $s$-wave Salpeter equation for the
generalized Hulth\'{e}n potential can be solved exactly. The relativistic
bound-state energy spectrum and the corresponding wave functions for the
Hulth\'{e}n potential have been obtained by the NU method. Some interesting
results including the ${\cal PT}{\rm -}$symmetric and pseudo-Hermitian
versions of the generalized Hulth\'{e}n potential have also been discussed
for the real bound-states. In addition, we have discussed the relation
between the non-relativistic and relativistic solutions and the possibility
of existence of bound states for complex parameters. Finally, we have shown
the possibility to obtain relativistic bound-states of complex quantum
mechanical formulations.

\acknowledgments S.M. Ikhdair wishes to dedicate this work to his
family for their love and assistance. This research was partially
supported by the Scientific and Technological Research Council of
Turkey.

\end{document}